\documentclass[preprint,aps]{revtex4}
\usepackage{amssymb}
\usepackage{amsmath}
\usepackage{epsf}
\usepackage{mathrsfs}
\usepackage[dvips]{graphics}
\usepackage[dvips,usenames]{color}
\begin{document}
\def \beq{\begin{equation}}
\def \eeq{\end{equation}}
\def \bea{\begin{eqnarray}}
\def \eea{\end{eqnarray}}
\def \bem{\begin{displaymath}}
\def \eem{\end{displaymath}}
\def \f{\frac{\Omega}{\omega}}
\def \fl{\Phi_{\Omega}}
\title{Composite bosons description of rapidly rotating 
Bose-Einstein condensates}
\author{Eric Akkermans and Sankalpa Ghosh\\
{\it Physics Department, Technion I.I.T. Haifa-32000, Israel}}
\date{\today}

\begin{abstract}
Recently a novel phase transition has 
been observed \cite{jalin} in a rotating
Bose-Einstein condensate
when the rotating
frequency $\Omega$ reaches the transverse trap frequency $\omega_{\bot}$
and eventually crosses it. We study certain aspects of this experiment 
in terms of the condensation of
composite bosons and the corresponding  vortex formation
using a Chern-Simon Gross-Pitaevskii theory.\\

\end{abstract}

\maketitle

\section{Introduction}
Following the observation of a regular array of vortices in a rotating
Bose-Einstein condensates (BEC) \cite{mit} there have been a great amount of
theoretical work \cite{wilkin1, wilkin2, paredes, lundh, ur1, ho, 
allan, tarun, japan, ur2, njp, ens, chen} and 
also some further experimental development \cite{exp1, exp2}
to understand the behaviour of the condensate when it is rotated
even faster. Apart from the fate of the vortex lattice
submitted to a faster rotation, the interest stems also from the
possibility of observing Laughlin-liquid like states in the condensate
when the rotational frequency becomes almost equal to the
transverse trap frequency ($\Omega \approx \omega_{\bot}$).
Since the number of atoms in the rotating condensates is very
large, it has been pointed out \cite{Fet1, Stringari1,
ur1} that one can still use the Thomas-Fermi (TF) approximation
where the kinetic energy is neglected compared to the interaction
energy, even though the system is very close to the Landau level limit.
Within the TF approximation it has been shown \cite{Fet1,ur1}
that the vortex lattice may survive when $\Omega$ is very close
to $\omega_{\bot}$. Ho \cite{ho} reached the same conclusion
using the Lowest-Landau level (LLL) approximation where  
the kinetic energy term is also frozen by the LLL constraint
and thereby it is equivalent in this respect to the
TF approximation.
In a subsequent development Sinova {\it et. al.} \cite{allan}
analyzed the
Bogoliubov spectrum in the LLL limit and reached the conclusion that
quantum fluctuations melt the vortex lattice. The same conclusion
was  reached by Wilkin {\it et. al.} \cite{wilkin2} using exact
particle diagonalization technique in a toroidal geometry.
Fishcher {\it et. al.} \cite{ur1} have addressed
the same problem using an effective free energy for
the interacting vortices in a rotating frame
and provided a  phase diagram
consisting of regions of Laughlin liquid of bosons, Laughlin
liquid of vortices and vortex lattice.

In a very recent experiment at ENS, Bretin
{\it et. al.}\cite{jalin} reached the Landau level (LL) limit
of a BEC. They superimpose an
additional quartic confining potential
above
the usual quadratic confining potential in the transverse plane,
thereby eliminating the possibility of centrifugal instability in
the limit  $\Omega=\omega_{\bot}$.
This experiment therefore reaches a regime where
the TF approximation does not hold anymore.
The vortex lattice is stable even at
a very high value of the rotational frequency
$\Omega$ (almost $98$ percent of $\omega_{\bot}$).
As $\Omega$ is further increased and eventually taken just beyond $\omega_{\bot}$, 
one obtains on both sides of $\Omega=\omega_{\bot}$
a blurred image of the vortex lattice
indicating a phase transition.  
As $\Omega$ is further increased beyond $\omega_{\bot}$
a condensate like
region reappears in a central region of the blurred image
and
 again new vortices start nucleating. At slightly higher $\Omega$, one
recovers
a collection of few such vortices again arranged in a regular array.

It should be pointed out that in most cases the rotational frequency
$\Omega$ is equal to the stirring frequency $\Omega_{stir}$. 
However, in the ENS 
experiment $\Omega_{stir}$ is not always equal to $\Omega$. 
Though in most cases 
the two are almost equal, it has been pointed out that beyond
$\Omega=\omega_{\bot}$ it is not correct anymore. 
Particularly at the highest
reached value of $\Omega_{stir}$ at which 
a vortex lattice becomes
clearly visible again, the  frequency $\Omega$ is predicted to be 
much lower than $\Omega_{stir}$.
However, it has been clearly demonstrated that when the rotational frequency 
of the condensate equals or exceeds $\omega_{\perp}$, the 
number of vortices in the condensate is much lower than expected 
at such a high $\Omega$.

In the present
paper we argue that those features are a manifestation of boson-vortex
duality
which has been earlier predicted
in the  framework of Chern-Simon (CS) theory
\cite{Jackiw, macgirv, read, duality, cslg, rajaraman, bqhe1, bqhe2}
in order to explain certain aspects of the
quantum Hall effect (QHE).
That such a composite boson can be formed at $\Omega=\omega_{\bot}$
has been first pointed out by Wilkin {\it et.al.} \cite{wilkin1}.
In the quantum Hall system, however, this
boson-vortex composites known as composite bosons remains a
useful theoretical tool
to understand the existence of the off-diagonal
long-range order (ODLRO), a property of the
one body density-matrix of the composite bosons rather than the one 
body density matrix of the electrons
\cite{macgirv, read, duality}.
However we shall argue that
for a rapidly rotating BEC, composite bosons
can actually be formed
since the vortices are the zeroes of the condensate wave function and
not a ficticious statistical flux. We propose to interpret  
the ENS experiment as a signature of the formation of composite-boson. 
We note that recently a CS field theory
has been proposed to describe  rotating bosons near a
Feshbach resonance where we also need to consider the formation of
molecules as a result of the strong interaction \cite{holland}.

This paper is organized as follows. We shall start with the energy
functional of the vortex lattice in the rotating frame as given by Fetter
\cite{Fet1} and show why it is energetically favourable to form at the frequency $\Omega
=\omega_{\bot}$ a
condensate of composite bosons
of almost uniform density in a region where the quartic
confinement can be considered to be smaller than the
other relevant energy scales. We then write
the free energy in terms of a Chern-Simon Gross-Pitaevskii
(CSGP)
theory which is very similar to
the Chern-Simon Landau-Ginzburg (CSLG) theory of the quantum Hall effect
\cite{cslg}.
We shall show how this expression of the free
energy can explain some of the new features observed in the ENS experiment\cite{jalin}.
We shall also point out that at the point $\Omega=\omega_{\bot}$ the
model is self-dual. Then we make some suggestions on
possible experiments to probe the nature of this new superfluid phase.
Finally we summarize our observations.

\section{Gross-Pitaevskii Energy Functional}
The ENS experiment is performed with $N=35. 10^{5}$~ $^{87}Rb$ atoms in a cylindrical
trap with $\omega_{\bot}=\omega_{x}=\omega_{y}=2\pi 65.6Hz, \omega_{z}
=2\pi11Hz$. In addition there is a confinement potential of the form
$kr^{4}$ with $k=6.545 10^{-12} S.I.$  \cite{jalin}.
For convenience, we shall first omit the quartic term assuming that
 it does not play any other role than making the condensate stable.
Nevertheless we must point out that this quartic potential on its own can add
interesting features as pointed out by Lundh \cite{lundh} (see also \cite{tarun1}) 
and besides it will influence the collective spectrum around
the mean-field solution we shall provide here.

We start with the Gross-Pitaevskii free energy of the bosons of mass 
$M$ in a rotating frame.
For the order parameter $\Psi$
\bea F &=&\int dV \big{[}
\frac{\hbar^2}{2M}(\boldsymbol{\nabla}\Psi)^2+V_{\bot}(r_{\bot})|\Psi|^2
\nonumber \\
& &\mbox{}+\frac{1}{2}M\omega_{z}^2z^2|\Psi|^2+\frac{1}{2}g|\Psi|^4
-\Psi^{\ast}\boldsymbol{\Omega}.\boldsymbol{r} \times \boldsymbol{p} 
\Psi \big{]} \label{energy}\eea
It has been pointed out
\cite{Fet1, Stringari1} that the vortex lattice experiences  a
solid body rotation \cite{feyn}
with a solid-body velocity $\boldsymbol{v_{sb}}$
given by

\beq \boldsymbol{\nabla} \times \boldsymbol{v_{sb}} = 2\boldsymbol{\Omega}
\eeq
The existence of this solid body rotation then ensures that neither the
phase $ \hbar S$ or the superfluid velocity $\boldsymbol{v_{s}}=\frac{\hbar}{M}
\boldsymbol{\nabla}S$ can be spatially periodic. The system thus behaves
locally like a superfluid in each unit cell of the lattice, but globally
as a rigid body provided the core size of each individual vortex is less than
a lattice constant.

Under these assumption the Gross-Pitaevskii energy functional can be written as \cite{Fet1}

\bea F &=&\int dV \big{[}
\frac{M}{2} (\boldsymbol{v_{s}}-\boldsymbol{v_{sb}})^2|\Psi|^2+
\frac{\hbar^2}{2M}(\boldsymbol{\nabla}\Psi)^2+V_{\bot}(r_{\bot})|\Psi|^2
\nonumber \\
&&\mbox{}+V_{cent}|\Psi|^2+\frac{1}{2}M\omega_{z}^2z^2|\Psi|^2
+\frac{1}{2}g|\Psi|^4
\big{]} \label{energy2}\eea
where $V_{cent}(r_{\bot})=-\frac{1}{2}M\Omega^2r_{\bot}^2$.

The Landau level limit is defined by the condition
$V_{\bot}(r_{\bot})=-V_{cent}(r_{\bot})$.

The minimum energy 
configuration, as we shall show using a singular gauge transformation leading to a
composites of bosons and vortices, is achieved in the mean-field approximation by 
considering $\boldsymbol{v_{s}}=\boldsymbol{v_{sb}}$ and 
$|\Psi|=constant$, namely when the kinetic energy of the condensate
in the presence of a vortex lattice is equal to zero. In
this limit the solid body rotation becomes identical to the
superfluid motion and we recover the superfluid
order parameter throughout the bulk. 

This new phase of composite
of bosons and vortices shows superfluidity, namely 
off-diagonal long range order (ODLRO) \cite{macgirv} for the
density matrix of the composite-bosons. It differs from
the ODLRO of the bare  atomic condensate with
vortices obtained at low rotating frequencies. We assume that the energy to 
bind bosons to vortices results from the repulsive interacting energy
of the original bosons, thus leading to a renormalization of $g$. This behaviour is  
very naturally described through a Chern-Simon Gross-Pitaevskii 
approach which has the advantage to be independent of the explicit 
LLL constraint.
For further discussion we also neglect the effect of confinement in the
$z$ direction, assuming that the profile of the condensate in this
direction is the same as for the non-rotating case.

\section{Chern-Simon Gross-Pitaevskii Theory}
Chern-Simon (CS) field theories have been originally proposed in order to
describe the behaviour of charged planar matter
interacting with photons whose dynamics is governed not only by the
usual Maxwell density $-\frac{1}{4}F^{\mu \nu}F_{\mu \nu}$
but also by the Chern Simon (CS) term
$\frac{\kappa}{4}\epsilon^{\mu \nu \sigma}F_{\mu \nu}a_{\sigma}$
which gives
rise to topologically massive $(2+1)$-dimensional electrodynamics 
\cite{Jackiw}. An important consequence  of the CS term is
that any charged excitation also carries a magnetic flux proportional to this charge.
For the low-energy,  long-wavelength physics
in $(2+1)$-dimensional electrodynamics the
CS Lagrangian is more important than the conventional
Maxwell Lagrangian since it
contains one less derivative. For a rapidly rotating BEC, the density 
of vortices is high and is assumed to form a vortex liquid, so that we 
can use the CS Lagrangian
in order to couple it to the bosonic matter density $|\Psi|^2$.
 
This coupling is achieved through the Chern-Simon transformation 
between the boson order parameter $\Psi$ and the order parameter $\Phi_{cb}$ 
describing the composite bosons, 
\beq \Phi_{cb}(\boldsymbol{r_1},\cdots,\boldsymbol{r}_N) =
e^{-im\sum_{i<j}\theta_{ij}}\Psi(\boldsymbol{r_1},\cdots,\boldsymbol{r}_N)
\eeq
where $\theta_{ij}$ is the angle describing the relative position of the 
$i$-th
and the $j$-th bosons. Here $m$
is always an even integer. In two spatial dimensions, this singular gauge transformation
attaches an integer number $m$ of angular momentum to each boson
so that

\beq \left [ \left ( \sum_{i}(-i\frac{\hbar}{M} \boldsymbol{\nabla}_{i}+ 
\boldsymbol{A}(\boldsymbol{r_{i}} \right ) \right] \Psi=
\left [ \left ( \sum_{i}(-i\frac{\hbar}{M} \boldsymbol{\nabla}_{i}+ 
(\boldsymbol{A}(\boldsymbol{r_{i}})+
\boldsymbol{a}(\boldsymbol{r_{i}})) \right ) \right ]\Phi_{cb} \eeq
where
\beq \boldsymbol{a_{i}}=\frac{\hbar m}{M}\sum_{i\neq j}\boldsymbol{\nabla}_{i}
\theta_{ij} \eeq
and $\boldsymbol{A}=\boldsymbol{\Omega}\times \boldsymbol{r}$.
Here indices refer to bosons and hence repeated indices does not mean
summation.
 In the CSLG description of QHE the flux of $\boldsymbol{a}$
is known as the statistical
flux. However here the vector potentials $\boldsymbol{a}$ are 
associated to the vortices already
present in the condensate, and in the limit
where the number of bosons is equal to the  $m$
times the number of the vortices
the bosons and vortices make composites.
One can physically view
this mechanism as the re-entering of bosonic matter in the vortices
which are zeroes of the condensate-wavefunction $\Psi$ and thereby making
the condensate irrotational again.
For $m=0$
one recovers the original Gross-Pitaevskii description.

The Chern Simon transformation in the second quantized language
is given by
\beq \hat{\Phi}_{cb}(\boldsymbol{r})
=e^{-\hat{J}(\boldsymbol{r})}\hat{\Psi}(\boldsymbol{r}) \eeq
where the operator $\hat{J}(\boldsymbol{r})$ is given by 
\beq \hat{J}(\boldsymbol{r})=im\int d\boldsymbol{r}
\hat{\Psi}(\boldsymbol{r})^{+}Im \log(z-
z')\hat{\Psi}(\boldsymbol{r}), z=re^{i\theta} \eeq
This tranformation at the mean-field level 
gives the phase of the Laughlin-wave function.
Considering fluctuations around the meanfield 
and neglecting higher order terms  
one can produce only the modulus of the 
Laughlin wavefunction \cite{cslg, bqhe1}. 
Using a modified form of the composite-boson field
operator $\hat{\Phi}_{cb}$ \cite{read}
(whose auto-correlation function shows true long-range behaviour),
Rajaraman and Sondhi \cite{rajaraman} later generalized  
this transformation to produce
the full Laughlin wavefunction 
at the mean-field level from  a second quantized
Chern-Simon description. The same procedure can be applied here also.

The Chern-Simon Gross-Pitaevskii Lagrangian of
the rapidly rotating bosons is written as

\beq L_{csgp}=L_{gp}+L_{cs} \label{lag}\eeq
with 
\beq L_{gp}=\int d\boldsymbol{r}
-\hat{\Phi}^{\ast}_{cb}(i\hbar \frac{\partial \hat{\Phi}_{cb}}{\partial t})+
\frac{\hbar^2}{2M}\hat{\Phi}^{\ast}_{cb}
\left (-i\boldsymbol{\nabla}-\frac{M}{\hbar}
(\boldsymbol{A}+\boldsymbol{a})\right )^2 \hat{\Phi}_{cb}+
g|\delta \hat{\rho}_{cb}|^2
\label{CSGP1}\eeq
where $\delta \hat{\rho}_{cb}=
\hat{\Phi}_{cb}^{\ast}\hat{\Phi}_{cb}-\overline{\rho}_{cb}$
and

\beq L_{cs}=\frac{M}{2\pi  \hbar m}
\int d\boldsymbol{r}
\epsilon^{\mu \nu \sigma}
a_{\mu}\partial_{\nu}a_{\sigma} \eeq

The term proportional to $\delta \hat{\rho}_{cb}$ 
accounts for the attractive interaction between composite bosons. It 
is mediated by the gauge-fields $\boldsymbol{a}$ and thus is long 
ranged unlike the interaction between
the bosons described by the order parameter $\Psi$. 

In the original problem of rotating bosons, 
the additional energy due to the applied rotation 
can be written as a sum of the self energy of each vortex 
and of an interaction term between the vortices which accounts for 
their short range and binary interacting potential \cite{agm}. The Chern Simon 
transformation maps this problem onto those of a liquid of interacting 
composite bosons where the dynamical gauge fields $\boldsymbol{a}$ 
play a role similar to the magnetic field in superconductors, by 
canceling the applied rotational velocity in the bulk of the system 
just like for the Meissner effect, leaving composite bosons 
whose interaction is described by the term proportional to $\delta \hat{\rho}_{cb}$ 
in (\ref{CSGP1}) where $\overline{\rho}_{cb}$ accounts for the fact that in the absence of 
applied rotation, the Lagrangian $L_{gp}$ vanishes at the mean field 
level. $L_{cs}$ is a Lagrange's
multiplier which implements the constraint relating the rotational
flux quanta to the original matter field density. Therefore, it does not 
contribute to the free energy which is identical to those given
in (\ref{energy}) when transformed back. 
To obtain the the mean-field solution we first 
replace the operators $\hat{\Phi}_{cb}, 
\hat{\Phi}_{cb}^{\ast}, \boldsymbol{a}$
in the Lagrangian (\ref{lag}) with the corresponding functions
$\Phi_{cb}(\boldsymbol{r}), \Phi_{cb}^{\ast}(\boldsymbol{r}), 
\boldsymbol{a}(\boldsymbol{r})$ 
by taking its expectation value in suitable field theoretical
state. The corresponding Euler-Lagrange equations of motion 
are obtained through the functional
derivatives with respect to  $\Phi_{cb}^{\ast}(\boldsymbol{r})$ and 
the dynamical gauge fields $\boldsymbol{a}(\boldsymbol{r})$~(For
further description we shall omit the argument $\boldsymbol{r}$).  
They are 

\bea \boldsymbol{\nabla} \times \boldsymbol{a} &=& -m\frac{h}{M}\rho_{cb}
\label{constraint} \\
\epsilon_{\alpha \beta}\left (\frac{\partial}{\partial t}a_{\beta} -
\frac{\partial}{\partial x_{\beta}}a_{0} \right )&=&-m\frac{h}{M}j_{\alpha}
 \label{ampere}\\
\frac{1}{2M}[-i\hbar \nabla - M(\boldsymbol{A}+\boldsymbol{a})]^2\Phi
_{cb} +\Phi_{cb}\delta \rho_{cb}
&=&0 \label{eqmotion} \eea

where  
\beq j_{\alpha}=\frac{\hbar}{2mi}
[\Phi^{\ast}_{cb}(\partial_{\alpha}\Phi_{cb}) 
-(\partial_{\alpha}\Phi^{\ast}_{cb})\Phi_{cb}] 
-(A_{\alpha}+a_{\alpha})\rho_{cb} \eeq
The first of these equations expresses the boson-vortex duality while
the second is the Maxwell Ampere's law. The last equation describes
the motion of the composite bosons under the combined effect of an 
applied rotation
and of the pseudo rotation generated by the vortices. The 
classical solution of these equations which minimizes
the  free energy is

\bea \Phi^{MF}_{cb}&=&\sqrt{\overline{\rho}_{cb}} \label{mf1}\\
\boldsymbol{A}+\boldsymbol{a}&=&0 \label{mf2}\\
        a_{0}&=&0 \label{mf3}\eea

This solution  corresponds to a constant composite
boson density and the complete cancellation of the external rotation
by the pseudo-rotation generated by the vortex.
This is equivalent
to the Meissner effect in type II superconductors and the field
$(\boldsymbol{\nabla} \times
\boldsymbol{a})$ plays here the role of the 
magnetic induction $\boldsymbol{B}$.
Such a description is possible only when the number of vortices is high
enough so that the discrete number of singular vortices can be replaced
by a continuous function representing vortex density. At very 
low temperature the system should be well  
represented by this  classical solution and small fluctuations 
around it. In a real system we expect to have  
density modulations and the corresponding gauge field
fluctuations.

What will be the effect of an increasing rotation? When $\boldsymbol{A}$
is changed from its meanfield value $-\boldsymbol{a}$, the additional
rotation gives a corresponding density modulation.
We shall now see how that will produce a vortex.
The vortex solution  has a  density profile which is identical
to the mean field solution at long distance.
Therefore one can write
\beq \Phi^{MF}_{cb}=\sqrt{\overline{\rho}_{cb}} e^{i \theta}
,~~ r \rightarrow \infty
\label{asm1} \eeq
In order for this solution to satisfy  aymptotically 
(\ref{eqmotion}),
\beq \boldsymbol{A}+\boldsymbol{a}=\frac{\hbar}{M r} \theta 
\label{asm2}\eeq
must be satisfied at $r \rightarrow \infty$. In this way the vortex
solution is accompanied by a change in the rotational velocity due to
the gauge field $\boldsymbol{a}$.
Upon integration, the right hand side of (\ref{asm2}) gives the extra flux ($\frac{h}{M}$) associated with the
deviation of $\boldsymbol{a}$ from its mean-field value $-\boldsymbol{A}$.
This is equal to the circulation flux associated with a single vortex.
Thus
vortices in composite-bosons are like quasi-hole excitations
over a  Laughlin-liquid like ground state \cite{paredes, cslg}.    
The
extra energy of a vortex can be evaluated by numerically
solving (\ref{eqmotion}) for a solution of the
form $f(r)e^{i\theta}$ which satisfies 
(\ref{asm1}) and  (\ref{asm2}). In the  CSLG theory of QHE such solutions
have been obtained by Tafelmayer, Curnoe and
Weiss\cite{vortex}. In the  bosons-vortices composites,
the energy required to create a vortex is going to be different 
from that  in a bare atomic condensate.
This is because of the renormalisation
in the interaction strength of the composite bosons.

The above description does not include the effect of the
confining quartic potential.
It leads to a more pronounced effect at the edge of the
system where the above description does not hold anymore.
It will also affect the collective
excitation spectrum around the mean-field solution.
It is possible that the chiral Luttinger liquid description
of QHE edge states
may capture the phenomenology at the edge \cite{wen}.

\section{Possible Probes for Composite Bosons}
The Chern-Simon description provides a relation between the 
composite-boson density and the circulation flux quanta \cite{duality, zee}. This explains qualitatively
why we should observe similar features
in the rotating composite-boson condensate and the rotating (bare)
Bose-Einstein condensate.
However, it has been already pointed out that
vortices in composite bosons are statistically
different from their counterpart in the boson condensate.
In the first
quantized language they are equivalent to quasihole
excitations in the bosonic Laughlin-liquid. A Berry phase
measurement like the one suggested by Paredes {\it et. al.}\cite{paredes}
may reveal their statistics. 

A composite-boson superfluid shows an ODLRO given in the mean field by
\beq G(z-z')=<\Phi^{MF}_{cb}|\hat{\Phi}_{cb}^{+}(z)\hat{\Phi}_{cb}(z')|
\Phi^{MF}_{cb}> \eeq
which can be observed in the
power law decay of the auto-correlation function of a single composite-boson \cite{discuss1}. 
In the QHE this correlation function is difficult
to measure since it is built out of a composite made of
real electrons and ficticious statistical fluxes. Here however, 
it may be even possible to measure
it directly  as the composites are made of
ordinary bosons and real vortices.

There is another important difference between the ordinary
BEC and the composite BEC.
We speculate that the interaction between the composite-bosons
is much weaker than between the original bosons. This
conjecture may be verified
by measuring the $s$-wave scattering length when these composite
bosons are formed. This may serve as an important step to probe the
composite boson superfluid.
Alternatively,
another possible way to probe the composite boson superfluid is to look
at the collective excitation spectrum \cite{lee}
and compare it with the Bogoliubov
spectrum of the bare condensate. The measurement of collective
excitations may also reveal the renormalisation of the interaction strength
between the composite bosons.

For $\Omega > \omega_{\bot}$ one can again
use the Thomas-Fermi approximation for the composite boson condensate, 
but with the renormalized interaction
strength. By increasing $\Omega$ further than 
what has been achieved in the ENS experiment, 
one may be able to obtain some information about this 
renormalized interaction strength 
from the evolution of the optical thickness of the atomic cloud
after a time of flight measurement and taking the $s$-wave scattering
length as a variational parameter in the resulting data
\cite{jalin}. 
In this limit, an increasing $\Omega$ gives rise to
an effective confinement potential 
of the form $(-ar^2+kr^4)$ with $a>0$ (Mexican hat)
and as
a result the size of the condensate region must grow. The ENS experiment
supports this observation.
In this
region the composite-bosons are again in the lowest Landau level and
their kinetic energy can be neglected compared to the other
terms in the expression of the
free energy. It will be interesting to study the
collective excitation spectrum using the Thomas-Fermi approximation in this regime
and to compare it to the  one obtained within the CSGP framework 
\cite{lee}.

At a frequency $\Omega=\omega_{\bot}$, the 
ENS experiment \cite{jalin} showed that the radial distribution of the cloud 
developed a shallow local minima
around the point $r=0$ where the mean-field CSGP approximation is better justified.
By reducing the strength $k$
of the quartic confinement, this region  will be enlarged. 
In that case one shall recover superfluidity and vortex
generation for $\Omega \ge \omega_{\bot}$ over a larger 
area compared to the present experiment.

\section{Summary and outlook}
We have proposed a Chern-Simon Gross-Pitaevskii theory to describe
the behaviour of a rapidly rotating condensate at $\Omega \ge \omega_{\bot}$.
The central idea in this formulation is the duality between
bosons and vortices.
We have studied certain features of
a recent experiment \cite{jalin} in terms of this theory.
The description in terms of composite bosons is valid in the 
limit where 
the number of vortices in the condensate  is 
of the same order as the number of atoms. 
There is no confirmation of this aspect
at least at the present stage of the experiment.
However this requires to probe the condensate in more details
around $\Omega=\omega_{\bot}$ through an adiabatic changing of the rotational frequency.
This should lead to a better comparison between the  theory
and the experiment. 
We also predict a renormalisation of the $s$-wave 
scattering length due to the formation of composite bosons.
The statistical phase-measurement type experiments already suggested
and the measurement of
collective excitations around $\Omega=\omega_{\bot}$
may give more informations about the condensate of composite-bosons, namely 
the ODLRO and the nucleation of vortices.

\vskip 20pt
\noindent{\large\bf Acknowledgements}\\\\
This work is supported by the Israel 
Council for Higher Education, the Technion, the  Israel Academy of Sciences 
and the fund for promotion of Research
at the Technion. We thank J. Dalibard 
for a very helpful correspondence and for making available
to us the experimental data of him and his group 
prior to the publication. We thank  
R. Rajaraman and Efrat Shimshoni for some useful correspondence.


\begin{thebibliography}{99}
\bibitem{jalin} V. Bretin, S. Stock, Y. Serin and J. Dalibard
{\it cond-mat}/0307464.

\bibitem{mit} J. R. Abo Shaeer, C. Raman, J.M. Vogels and W. Ketterle,
Science, {\bf 292}, 476 (2001)


\bibitem{wilkin1}N. Wilkin and J.M.F. Gunn, Phys. Rev. Lett, {\bf 84}, 6
(2000)

\bibitem{wilkin2}N. R. Cooper, N. Wilkin and J.M.F. Gunn, Phys. Rev. Lett,
{\bf 87}, 120405(2001)

\bibitem{paredes}B. Paredes, P. Fedichev, J. I. Cirac, P. Zoller,
Phys. Rev. Lett, {\bf 87}, 010402 (2001)
\bibitem{lundh}E. Lundh, Phys. Rev. A.,{\bf 65}, 043604, (2002)
\bibitem{ur1}U.R. Fischer and G. Baym, Phys Rev. Lett, {\bf 90}, 140402
(2003)
\bibitem{ho}T. L. Ho, Phys. Rev. Lett, {\bf 87}, 060403(2001),
\bibitem{allan} J. Sinova, C.B. Hanna and A.H. MacDonald, Phys. Rev. Lett,
{\bf 89}, 034403(2002)
\bibitem{tarun}T.K.Ghosh and G. Baskaran, {\it eprint} {\it cond-mat}-0207484
\bibitem{japan}K. Kasamatsu, M. Tsubota, M. Ueda, Phys. Rev. A, {\bf 66}, 
053606(2002)
\bibitem{ur2} U. R. Fischer, P. O. Fedichev, A. Recati, P Zoller
{\it cond-mat}/0212419
\bibitem{njp}G.M.Kavoulakis and G. Baym, New Jour. of Physics. {\bf 5}, 51.1
(2003)
\bibitem{ens}N. Regnault and Th. Jolicoeur, Phys. Rev. Lett, {\bf 91}, 
030402, (2003) 
\bibitem{chen}Z.B.Chen, B. Zhao, Y. D. Zhang,{\it cond-mat}/0211187
\bibitem{exp1}P.Rosenbusch {\it et.al.}, Phys. Rev. Lett. {\bf 88}, 
250403(2002)
\bibitem{exp2}P. Engels. {\it et. al.}, Phys. Rev. Lett. {\bf 90}, 
170405 (2003)
\bibitem{Fet1} A.L.Fetter, Phys. Rev.A., {\bf 64}, 063608(2001)

\bibitem{Stringari1}
M.Cozzini and S. Stringari, Phys. Rev.A.,{\bf 67}, 041602(R)(2003)
\bibitem{feyn}R. Feynman in {\it Progress in Low. Temp. Phys.}
edited by C.J.Gorter, North Holland Pub. Co., Amsterdam (1955)
\bibitem{tarun1}T.K.Ghosh, {\it cond-mat}/0306563

\bibitem{Jackiw}R.Jackiw and S.Y.Pi, Phys. Rev. D.,{\bf 42},3500(1990),
R. Jackiw and E.J.Weinberg, Phys. Rev. Lett, {\bf 64}, 2234(1990)

\bibitem{macgirv}S.M.Girvin and A.H. MacDonald, Phys. Rev. Lett, {\bf 58},
1252(1987)

\bibitem{read}N.Read, Phys. Rev. Lett, {\bf 62}, 86(1989)
\bibitem{cslg}S.C.Zhang, T.H. Hansson and S. Kivelson, Phys. Rev. Lett.
{\bf 62}, 82(1989),
S.C. Zhang, Int. J. of Mod. Phys. B {\bf 6}, 25(1992)
\bibitem{duality} D.H.Lee and M.P.A.Fisher, Int J. of Mod. Phys. B, {\bf 16
and 17}, 2675 (1991)
\bibitem{holland}S.G. Bhongale, J.N. Milstein and M.J. Holland, {\it cond-mat}/0305399
\bibitem{zee} {\it Quantum Field Theory in a Nutshell} by A. Zee
, Princeton University Press (2003)
\bibitem{rajaraman}R. Rajaraman and S.L.Sondhi, Int. J. of Mod. Phys. B
{\bf 7},793, (1996)
\bibitem{agm} E.Akkermans, D. Gangardt and K. Mallick, Phys. Rev.{\bf B62},12427-12439 (2000)
\bibitem{bqhe1} {\it The Quantum Hall Effect}
by D. Yoshioka, (Springer, 2002)
\bibitem{bqhe2}{\it Quantum Hall Effects: Field Theoretical Approach and
Related Topics} by Z. F. Ezawa, (World Scientific, Singapore, 2000)
\bibitem{vortex}R. Tafelmayer, Nucl. Phys. B {\bf 396}, 386(1993)
S.Curnoe and N. Weiss, Int J. Mod. Phys. A, {\bf 11}, 329(1996)
\bibitem{wen}X.G.Wen, Int J. of Modern Phys. B, {\bf 6}, 1711(1992),
 M.A. Cazalilla, {\it cond-mat}/0207715.
\bibitem{discuss1} The ODLRO defined here decays algebraically and
hence shows quasi-long range order. A modified ODLRO can be defined \cite{read} 
which shows true long-range order. The corresponding
CSLG which involves a non-unitary transformation between composite
bosons and bosons was developed in ref. \cite{rajaraman}.

\bibitem{lee}D.H.Lee and S.C.Zhang, Phys. Rev. Lett., {\bf 66}, 1220(1991)
\end{thebibliography}
\end{document}